\documentclass[letterpaper,twocolumn,10pt]{article}
\usepackage{usenix2019_v3}

\usepackage{tikz}
\usepackage{amsmath}

\usepackage{filecontents}

\usepackage{xspace}

\usepackage{verbatim}

\usepackage[]{hyperref}

\usepackage{pifont}

\usepackage{makecell}

\usepackage{adjustbox}

\usepackage{graphicx}
\usepackage{subcaption}

\usepackage{multirow}
\usepackage{multicol}

\usepackage[table]{xcolor}

\usepackage{rotating} 

\usepackage{booktabs}

\usepackage{algorithm}
\usepackage{algpseudocode}
\usepackage{titlesec}

\usepackage{enumitem}

\usepackage{soul}

\usepackage{tcolorbox}

\usepackage{mdframed}

\begin{document}

\date{}

\title{Learning From Developers: Towards Reliable Patch Validation at Scale for Linux}

%
%


\author{
{Chih-En Lin} \\
Purdue University
\and
{Attreyee Mukherjee} \\
Purdue University
\and
{Ajay Rawat} \\
Purdue University
\and
{Ruqi Zhang} \\
Purdue University
\and
{Pedro Fonseca} \\
Purdue University
}

\maketitle


\def\sys{\textsc{FLINT}\xspace}
\def\llmonly{{\textsc{FLINT}$_{\text{LLM}}$}\xspace}
\def\kernel{{Linux}\xspace}
\def\version{{v6.15-rc5}\xspace}
\def\issue{{issue}\xspace}
\def\issues{{issues}\xspace}
\def\lkml{{LKML}\xspace}
\def\problem{{buggy patch}\xspace}
\def\solution{{bug-fix patch}\xspace}

\def\geminitfpro{{Gemini-2.5-Pro}\xspace}
\def\geminitfflash{{Gemini-2.5-Flash}\xspace}
\def\deepseekroeb{{DeepSeek-R1:32b}\xspace}
\def\qwentttb{{Qwen3-coder:30b}\xspace}
\def\gemmatfb{{Gemma3:4b}\xspace}
\def\gemmatttb{{Gemma3:12b}\xspace}

\definecolor{darkgreen}{rgb}{0.0, 0.5, 0.0}

\def\Snospace~{\S{}}
\renewcommand*\sectionautorefname{\Snospace}
\def\sectionautorefname{\Snospace}
\def\subsectionautorefname{\Snospace}
\def\subsubsectionautorefname{\Snospace}


\newcommand\para[1]{\vspace{2pt} \noindent \textbf{#1}}

\newcommand{\takeaway}[2]{
    \begin{mdframed}[skipabove=10pt, skipbelow=10pt] 
        \noindent \textbf{Finding:} #1
        
        \vspace{5pt} 
        \noindent \textbf{Implication:} #2
    \end{mdframed}
}

\titlespacing*{\section}     {0pt}{1.0ex plus 1ex minus .2ex}{0.75ex plus .2ex}
\titlespacing*{\subsection}   {0pt}{1.0ex plus 1ex minus .2ex}{0.1ex plus .2ex}
\titlespacing*{\subsubsection}{0pt}{1.0ex plus 1ex minus .2ex}{0.1ex plus .2ex}

\sloppy

\subsection*{Abstract}

Patch reviewing is critical for software development, especially in distributed open-source development, which highly depends on voluntary work, such as \kernel.
This paper studies the past 10 years of patch reviews of the \kernel memory management subsystem to characterize the challenges involved in patch reviewing at scale.
Our study reveals that the review process is still primarily reliant on human effort despite a wide-range of automatic checking tools.
Although kernel developers strive to review all patch proposals, they struggle to keep up with the increasing volume of submissions and depend significantly on a few developers for these reviews.

To help scale the patch review process, we introduce \sys,
a patch validation system framework that synthesizes insights from past discussions among developers and automatically analyzes patch proposals for compliance.
\sys employs a rule-based analysis informed by past discussions among developers and an LLM that does not require training or fine-tuning on new data, and can continuously improve with minimum human effort.
\sys uses a multi-stage approach to efficiently distill the essential information from past discussions.
Later, when a patch proposal needs review, \sys retrieves the relevant validation rules for validation and generates a reference-backed report that developers can easily interpret and validate. 
\sys targets bugs that traditional tools find hard to detect, ranging from maintainability issues, e.g., design choices and naming conventions, to complex concurrency issues, e.g., deadlocks and data races.
\sys detected 2 new issues in \kernel v6.18 development cycle and 7 issues in previous versions.
\sys achieves 21\% and 14\% of higher ground-truth coverage on concurrency bugs than the baseline with LLM only.
Moreover, \sys achieves a 35\% false positive rate, which is lower than the baseline.
\section{Introduction}
\label{sec:intro}

Large and widely-used open source software systems, particularly for systems like \kernel kernel, rely on distributed development to achieve high code quality.
However, maintaining high quality requires extensive code review processes, which are unreliable and challenging to scale~\cite{corbet25:state-mm-dev, corbet16:linux-dev-study} are complex and require extensive domain knowledge.

The \kernel kernel community uses various tools to enhance code robustness.
For example, static analysis tools~\cite{smatch, sparse-doc, gcc-static-analysis} are used to identify semantic issues and specific bug patterns.
Additionally, dynamic tools, such as fuzzers~\cite{syzbot, syzkaller, yang25:ml-fuzz-kernelgpt, gong23:ml-fuzz-snowcat, gong25:ml-fuzz-snowplow, wang21:ml-fuzz-syzvegas} and other automated testing tools~\cite{intel-lkp, ltp}, are executed around the clock to detect a range of runtime errors in Linux.
Although these tools are useful, there are classes of problems they cannot easily check, especially in a very dynamic codebase.
This includes providing feedback on design trade-offs and choices, identifying complex bugs caused by subtleties of the complex kernel API, and addressing non-compliance with code conventions.
As a result, during the patch review, developers still have to do a lot of the work, requiring manual effort that could have been used in other tasks, creating development bottlenecks, and making the process difficult to scale.
Our analysis (\autoref{sec:motivation:10-years}) shows that only $7.3\%$ of reviews are based on tool checks, whereas $92.7\%$ of the reviews are from developers' manual analysis.

\begin{figure}[t]
    \centering
    \includegraphics[width=0.475\textwidth]{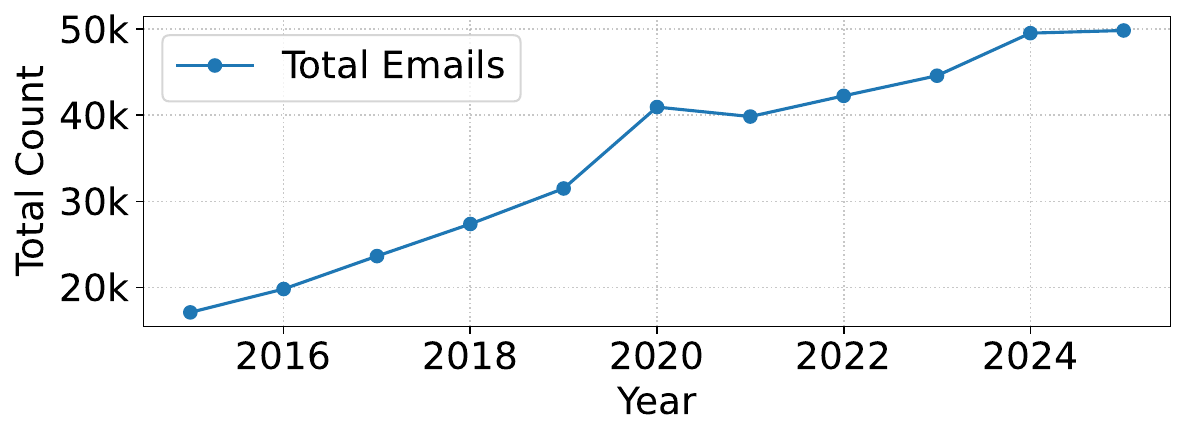}
    \caption{Emails exchanged in the Linux memory management mailing list from 2015 to November 2025.}
    \label{fig:nr-emails-10years}
\end{figure}

The non-scalable patch review process has recently emerged as a major bottleneck in the Linux development~\cite{corbet25:state-mm-dev}.
Recently, the gap between the growth of patch proposals and the number of developers available to review them has become a problem.
Our analysis of the \kernel's memory management, a core and the most active Linux subsystem, is illustrative of the severity of this problem.
Over the past 10 years, the number of emails exchanged in the memory management subsystem list has increased steadily from 17,108 in 2015 to 49,855 in 2025, as shown in \autoref{fig:nr-emails-10years}.
This activity led to impressive advances in the kernel -- the memory management subsystem now achieves much higher scalability and performance while also supporting a wide range of new features that handle application needs.
However, this also made the memory management subsystem grow by more than 5 times, leading to the adoption of new and complex system designs.
In fact, the review period -- from patch proposal to completion -- often exceeds one year~\cite{corbet25:state-mm-dev}.
This lengthy review process is largely due to the complex modifications involved in many patch proposals and limited expertise for reviewing them~\cite{corbet25:state-mm-dev, corbet16:linux-dev-study, abal14:variability-bug-study-linux}.
Moreover, it was found that the patches, which were merged before kernel version 5.8, introduced more bugs than were fixed, leading to the introduction of poor quality code, degrading the kernel~\cite{corbet25:dev-stat-616, corbet25:dev-stat-617}.

To improve the reliability and effectiveness of the review process, some works~\cite{sun25:llm-code-review, corbet25:llm-for-patch-review, google-sashiko} have used large language models (LLMs) to identify bug patterns in past data and systematically apply these patterns to detect bugs in new codebases or patches.
However, these approaches either have an approximately 50\% false positive rate of reporting bugs~\cite{corbet25:llm-for-patch-review}, detect simple bugs on less mature subsystems like drivers~\cite{yang25:knighter-transforming-static-analysis}, or require human effort~\cite{sun25:llm-code-review} to annotate the feedback for fine-tuning or training the model to keep pace with software updates.
As a result, there is a critical need for scalable patch review tools capable of identifying the conventional, complex, and high-priority errors that human reviewers target.

To address this need, this paper analyzes the root causes of the patch review process inefficiency by analyzing data from the past 10 years in the Linux Kernel Mailing List (\lkml).
Our study reveals that, despite the community's efforts to maintain a steady pace in reviewing the volume of patch proposals submitted to the \lkml, 73.87\% of them have not been reviewed in the past 10 years.
We estimate that approximately 21.5\% of patches require maintenance work, such as merging them into the source tree.
Additionally, 92.7\% of the review feedback comes from human reviewers, and 51.8\% of that feedback pertains to design and style aspects, including implementation decisions, coding style, and documentation.
This suggests a critical need for an automatic and reliable patch validation system that can support the kernel.

This paper introduces \sys, a rule-based, interpretable, LLM-modularized patch validation system.
\sys does not require training or fine-tuning the model and uses an attached rule set that can be collected automatically.
\sys aims to provide an automated system that triage the majority of issues and offers actionable and interpretable feedback for developers to fix them.
Ultimately, our goal is not to replace maintainers, but instead to provide a useful signal to developers proposing patches that can relieve the significant workload kernel maintainers face, and the patch proposals are still validated by a human before acceptance.

Given a developer discussion, \sys automatically analyzes the discussion and summarizes the critical concepts discussed into rules.
These rules provide reliable and consistent content that leverages the powerful reasoning capabilities of LLMs while minimizing the risk of LLM hallucination, resulting in a reduced false positive rate.
Moreover, \sys provides bug reports with the rules used and their provenance to maintain interpretability, keeping the developer in control of the process.
\sys uses these rules and the kernel source code to validate patch quality from various perspectives.
This includes detecting issues related to function naming, code conventions, and race conditions, capabilities that remain difficult for traditional tools to match.

Two main challenges need to be addressed to implement and evaluate \sys.
First, how to design the consolidated rule set to provide the comprehensive concepts from the large number of past discussions while maintaining the LLM performance.
Specifically, a previous study~\cite{hong25:context-affect-llm-perf} shows that a larger size of the context window can affect the LLM performance significantly.
\sys solves this problem by processing past discussions in multiple stages to extract the concept.
Whenever \sys extracts the rules from a single discussion thread, \sys generates raw rules and applies a filtering process to remove duplicates and non-substantive rules. 
After filtering, \sys categorizes the rules into two types, code logic and code convention, to have a more fine-grained dataset that can continually improve quality.
Then, \sys will merge the new rules into the existing rule set with some conditions to preserve detailed information while reducing the context window size.

The second challenge is to measure the accuracy and quality of issues reported by \sys in real-world scenarios.
Traditional machine learning metrics, such as F1 score and recall, are inapplicable due to the lack of an established ground truth.
The reported \issues fall into three types: the fixed bugs, undiscovered bugs, and non-bugs.
However, because the space of potential candidates, including false negatives in fixed and undiscovered bugs, as well as true negatives in non-bugs, is effectively infinite, it is impossible to enumerate all the possibilities for classification.
To address this problem, we propose a Ground-truth Coverage Score (GCS), which references the metrics from reinforcement learning rewards~\cite{zhang25:rl-generative-verifiers, zhao25:learning-reason-external-rewards, zhou25:reinforcing-general-reasoning-verifiers, gunjal25:rubrics-rewards-rl, jia25:writing-zero-bridge-gap-non-verifiable, huang25:reinforcement-learning-rubric-anchors}.
First, GCS instructs a target system, such as \sys, to generate the \issues based on the specific \problem.
Second, GCS employs the LLM-based verifier~\cite{cobbe21:training-verifiers-solve-math, lightman23:lets-verify-step-by-step, wang24:math-shepherd-verify-reinforce-llms, zheng23:judging-llma-with-mtbench} using Chain-of-Thought (CoT) reasoning~\cite{wei22:chain-of-thought-prompting} to quantify the extent to which the generated \issues cover the concepts in the ground truth, \solution.
To avoid hallucination, GCS uses majority voting~\cite{wang23:self-consistency-improves} and best-of-N~\cite{charniak05:coarse-to-fine-n-best, cobbe21:training-verifiers-solve-math} to avoid the inconsistent LLM behavior.

Our evaluation shows that our CoT-verifiers can achieve close to human performance in judging \issues that are close to the ground truth.
The evaluation also shows the performance benefits of \sys for the multiple-stage rule and issue generation with several benchmarks.
\sys can identify complex bugs, such as data races, fragmented locking, and deadlocks.
Moreover, \sys detects 2 new issues in \kernel v6.18 development cycle, and 7 issues in past versions.
\sys improves ground truth coverage by 21\% over the baseline with LLM only.
\sys provides consistent \issues, and performs 1.6 times lower false positive rate than the baseline. 

In summary, this paper makes the following 
contributions:
(1) a quantitative analysis of \kernel's memory management subsystem patch review process 
(2) the design and implementation of \sys, a first patch validation system for \kernel; and
(3) a comprehensive metric framework of the patch validation system.
\section{Motivation}
\label{sec:motivation}

In this section, we analyze the past 10 years of patch reviews for \kernel on the Linux Kernel Mailing List (\lkml) to identify the challenges associated with the existing tools.
First, we explain how we selected the dataset.
Next, we discuss the analysis of \kernel code reviews and evaluate the benefits and drawbacks of existing approaches to patch validation.

\subsection{Methodology of Study}
\label{sec:motivation:methodology}

\para{Thread Selection.}
We select the memory management subsystem from the \lkml as the data source, as it is one of the core, mature subsystem in \kernel with substantial developer participation.
Moreover, the memory management subsystem often intersects with several crucial system topics, resulting in other subsystems, such as the file system, driver, and hardware, frequently submitting their patches to this subsystem.
Therefore, using the memory management subsystem as the source ensures broad coverage of kernel activities while maintaining a manageable workload for analysis.
We collected patch discussion threads over the past 10 years, from 2015 to November 2025.
The trend in total email volume is shown in \autoref{fig:nr-emails-10years}.

\para{Random Sampling.}
The memory management subsystem contains $386,535$ messages over the past 10 years.
Given this scale, a detailed manual analysis of the entire dataset is infeasible.
Therefore, we generated a randomized list of message IDs for each year to conduct a manual analysis.
We then iterated through these randomized lists to draw a representative sample.

\subsection{10 Years Patch Review Analysis}
\label{sec:motivation:10-years}

\begin{figure}
    \centering
    \includegraphics[width=0.475\textwidth]{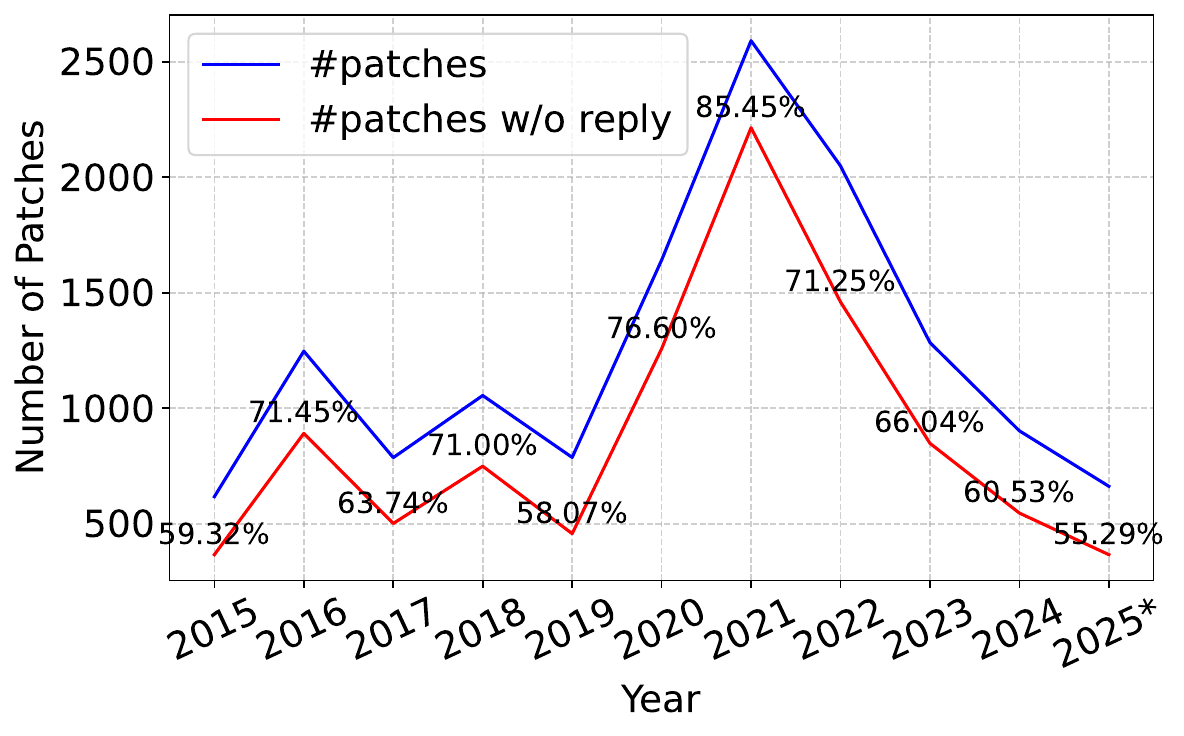}
    \caption{Number of patches and without any reply in the mailing list of the memory management subsystem from 2015 to Nov 2025, with random sampling of the 100 threads for each year. The number in the figure for each year is the percentage of the patches without any reply.}
    \label{fig:nr-patch-wo-reply}
\end{figure}

\para{Unreviewed Patches.}
To determine whether the review workload of developers and maintainers can handle the increased throughput of the emails in the mailing list, we analyze the percentage of patches that have not received any replies until November 2025. 
We randomly sampled 100 discussion threads for each year.
As shown in \autoref{fig:nr-patch-wo-reply}, the percentage of patches without any replies has consistently exceeded $50\%$ throughout 10 years in our dataset.
This suggests that, despite an increase in the volume of emails and patches submitted to the mailing list, both maintainers and developers have intensified their efforts in code review over the past decade.
However, the workload associated with code reviews has not been adequate to handle the large number of patches submitted.

\takeaway{
Over the past 10 years, the percentage of patches without any replies has consistently exceeded $50\%$.
This indicates that the code review process is fundamentally unscalable.
}{
The inherent scalability limits of patch review necessitate the adoption of automated tooling to assist developers in prioritizing and triaging the overwhelming volume of incoming contributions.
}

\para{Type of Review Feedback.}
To further understand where developers and maintainers spend most time reviewing, we examine the reply emails in the patch series.
We randomly sampled 20 reply emails for each year.
First, we manually categorize the reply email into three major categories:
(1) static tools~\cite{smatch, sparse-doc, gcc-static-analysis}, like GCC static analysis and kernel static analysis tools;
(2) dynamic tools~\cite{syzkaller, syzbot, intel-lkp}, such as fuzzer and runtime testing;
(3) and human, such as kernel developers and maintainers.
Second, we break down the human category into four subtypes, design, security, style, and others, to investigate what kind of feedback is most frequently provided by the reviewers.
The design type includes content that primarily discusses decisions regarding the logic and design of implementations, such as tradeoffs and corner cases.
The security type contains the content that mainly discusses the bug or security concerns.
The style type covers the feedback related to documentation, coding conventions, such as function types, and coding style.
We classify the rest of the reply emails as others if the email only contains feedback related to signatures and email communications, such as CCing someone who was missed or casual comments unrelated to the patches.
Finally, if one reply email has more than one type, we choose the most dominant type, defined as the primary focus or most important aspect of the discussion.
For example, if a reviewer mentions variable naming and design choice simultaneously~\cite{hillf15:patch-prior-category-example}, we categorize this email response within the design category.

\begin{figure}
    \centering
    \includegraphics[width=0.475\textwidth]{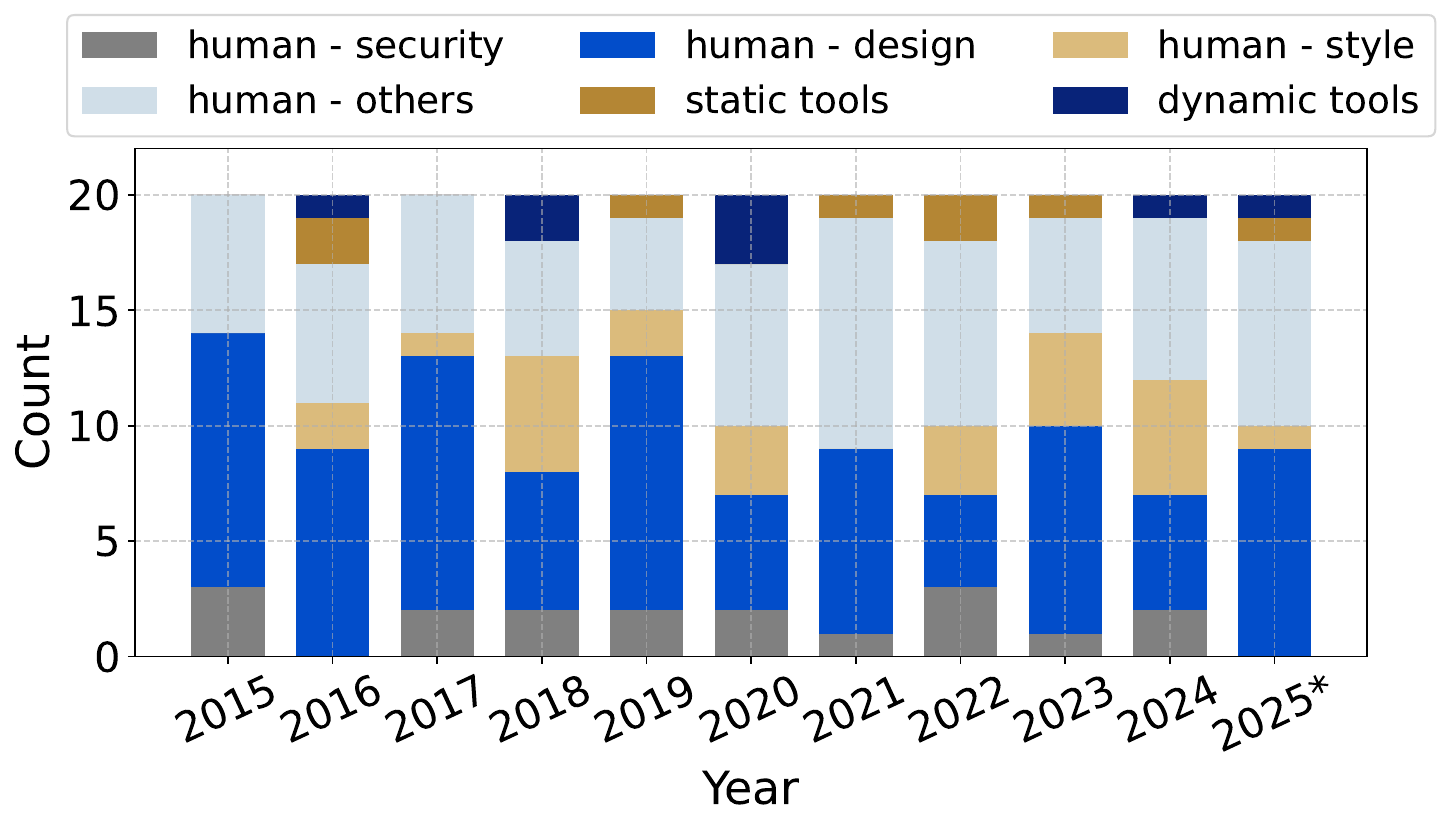}
    \caption{Distribution of the feedback came from. The dataset is a random sampling of 20 reply emails for each year.}
    \label{fig:reply-email-types}
\end{figure}

As shown in \autoref{fig:reply-email-types}, the human feedback dominates the review process ($92.7\%$), while the tools only contribute to a minority of code reviews ($7.3\%$).
In the human category, excluding the other type ($32.7\%$), design feedback is the most prevalent ($40.0\%$), followed by style ($11.8\%$) and security ($8.2\%$).
This design feedback often involves specific knowledge of the subsystem.
For example, one reply clarifies that \texttt{smp\_mb\_\_after\_atomic()} can only be used with the read-modify-write (RMW) atomic operations rather than non-RMW atomic operations, e.g., \texttt{atomic\_read()}, should use \texttt{smp\_mb()} or other primitives.

\takeaway{
Human experts carry the primary burden of the review process ($60\%$), primarily addressing complex design flaws ($40\%$) rather than simple stylistic errors.
}{
Current automation is insufficient because it targets low-level compliance.
To meaningfully reduce reviewer load, tools must evolve to understand domain-specific semantics and architectural logic.
}

\subsection{Patch Review Analysis}
\label{sec:motivation:historical}

We conduct a thorough analysis of the workload details to identify the specific types of challenges that kernel developers face.
We focus on the 2024 dataset, the most recent and most active period from the past 10 years, providing a comprehensive overview of the developers' workload throughout the year.
We also perform an automated analysis of data from 2022 and 2023, but the results do not show significant changes from 2024.

\para{Patch Thread Selection.}
We randomly sample 20 threads from each month in 2024, removing duplicates to avoid double-counting threads that span across consecutive months.
As shown in \autoref{tab:lkml_2024_analysis}, our dataset consists of $5,717$ emails from $228$ threads.
$1,633$ of these emails are patch proposals submitted for reviews, $3,950$ are emails replied to them, and the rest are emails unrelated to patch reviews, such as conference proposals.

\begin{table}[t]
\footnotesize
\centering
\begin{minipage}[t]{0.18\linewidth}
    \centering
    \begin{tabular}{lr}
        \toprule
        \textbf{} & \textbf{Total} \\
        \midrule
        Threads & 228 \\
        Emails & 5,717 \\
        Replies & 3,950 \\
        Patches & 1,633 \\
        \bottomrule
    \end{tabular}
\end{minipage}%
\hspace{0.20\linewidth}%
\begin{minipage}[t]{0.60\linewidth}
    \centering
    \begin{tabular}{lr}
        \toprule
        \textbf{} & \textbf{Perc.} \\
        \midrule
        Patches requiring maintenance & 21.5\% \\
        Patches without review & 53.2\% \\
        Patches with maintainer review & 23.2\% \\
        Maintainer email reply rate & 36.4\% \\
        \bottomrule
    \end{tabular}
\end{minipage}
\caption{\lkml 2024 sampled dataset properties.}
\label{tab:lkml_2024_analysis}
\end{table}

\para{Patch Thread Property.}
We analyze the basic properties of each patch thread to understand how much effort developers and maintainers spend on code review.
First, in terms of response latency, the results show that $69.7\%$ of threads receive an initial reply within 5 hours, and $35.1\%$ within just 2 hours. 
However, the length of the discussion remains long.
Among threads that receive a response, the median length is nearly 11 replies.
This shows that while the memory management community is responsive, the review process demands substantial and sustained effort from developers and maintainers.

\para{Maintenance Work.}
We manually analyze the tasks that are exclusive to maintainers, the work that cannot be distributed to non-maintainer developers, who are the vast majority of the reviewers.
This work includes tasks such as maintaining patches and the source tree, like merging and dropping patches.
The results show that only $21.5\%$ of the patches require maintenance work.
This suggests that most of the maintainers' workload does not primarily stem from maintenance tasks.

\para{Patch Response.}
We analyze the number of patch proposals that remain unreviewed and determine how many emails receive responses from maintainers.
This helps us to understand if the current code review process adequately covers all patch proposals and whether maintainers are engaging with the emails.
We find that there are $53.2\%$ patch proposals that have not been reviewed, and maintainers responded to only 23.2\% of the total patch proposals.
Moreover, we found that the code review workload is significantly unbalanced among maintainers.
Although the maintainer email reply rate, i.e., the percentage of total reply emails sent by maintainers, is 36.4\%, impressively, a single maintainer is responsible for 13.7\% of these responses.
One of the potential reasons~\cite{corbet25:state-mm-dev} is that most developers are not confident enough to review complex patch proposals, specifically, those outside of their comfort zone.

\takeaway{
While the community is highly responsive ($69.7\%$ of replies are sent in less than $5$ hours) and administrative overhead is low ($21.5\%$), over $53.2\%$ of patch proposals remain without a review, and the workload is heavily skewed toward a few experts. 
}
{
The bottleneck of patch review is the lack of experts.
The current reliance on a few maintainers to sustain deep technical discussions is insufficient to cover the volume of complex patch proposals, demanding tools that augment reviewer capacity.
}

\subsection{Towards Scaling Linux Development}

Although there are several tools available to find bugs and formatting issues (\autoref{sec:related-work}), \kernel is highly dynamic, and the existing tools are not able to detect the issues found in the vast majority of patch proposals.
Many existing tools rely on static analysis with high false positives, require version-specific kernel annotations, or ultimately have a narrow scope of checks.
The need to accommodate the dramatic and ongoing increase in the workload of Linux maintainers motivates us to explore automated techniques that can address an exceptionally large, complex, and dynamic codebase.
\section{Design}
\label{sec:design}

This section discusses the design of \sys and shows how to achieve summarization of the past discussion threads to create reliable rules, use the rules to analyze the patch proposal, and report the \issues while reducing the false positives of \issues.
We first explain the \sys system model.
Then, we present the multiple-stage rule extraction and how we extract the important information from past discussions and synthesize the rules.
Next, we discuss how the multiple-stage patch validation in \sys works to synthesize the reliable \issues.

\begin{figure}
    \centering
    \includegraphics[width=0.475\textwidth]{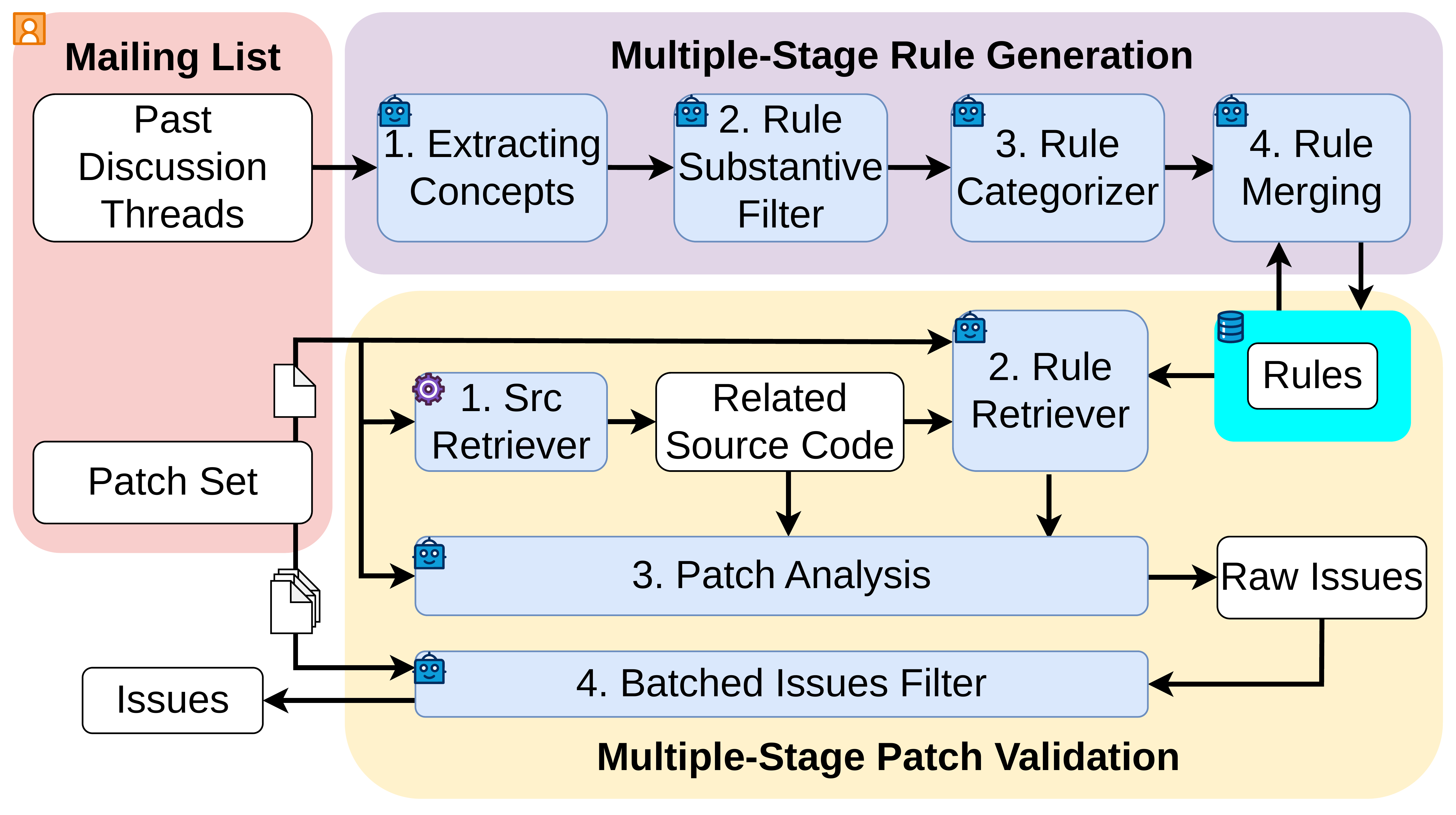}
    \caption{System Overview.
    }
    \label{fig:sys-overview}
\end{figure}

\subsection{System Model}
\label{sec:design:system-model}

\autoref{fig:sys-overview} shows the system model of \sys.
\sys extracts the concepts from past discussion threads through multiple stages to summarize into rules.
These stages ensure a broad coverage of the concepts while distilling the important information, fitting within the LLM context window, and then summarizing discussions on similar topics into individual rules.
After generating the rule set from past developer discussions, \sys validates the patch proposal that includes the cover letter, the commit message and the code, in incoming patches using the rule set.
\sys uses multiple stages of validation to retrieve relevant rules and source code, synthesize the \issues, and filter out the false \issues.

\para{Multiple-Stage Processing.}
\sys processes data in multiple stages for both rule generation and patch validation to decompose a complicated task into a set of simplified tasks.
This allows LLMs to focus on one specific goal, ensuring high performance~\cite{chen24:llm-perf-of-hard-easy-task-filter-vote}.
Furthermore, this approach reduces the size of the LLM context window, mitigating the performance degradation often caused by excessive context length~\cite{hong25:context-affect-llm-perf}.

\begin{figure}[t!]
    \centering

\begin{tcolorbox}[
    title={Example Rule},
    colback=black!5!white, 
    colframe=black!75!white, 
    coltitle=white, 
    fonttitle=\bfseries
]
\small

\para{Content:} In critical kernel paths that must guarantee forward progress, such as I/O, memory reclamation, or filesystem writeback, do not introduce potentially failing memory allocations without implementing a robust fallback mechanism. Instead of causing a hard failure or panic, the system must be able to continue operating, for instance by using pre-allocated resource pools, limiting I/O depth, or using reserves.



\para{Sources}\\
- message-id: <source of the message ID> \\
- author: <the author of source>

\para{History} \\
- <previous rule content> \\
- <previous rule content>

\end{tcolorbox}
\caption{A simplified rule generated by \sys.}
\label{fig:example-rule}

\end{figure}

\subsection{Rule Generation}
\label{sec:design:rule-generation}

In this section, we discuss the multi-staged pipeline in \sys for transforming past discussions into a consolidated rule set.
We first describe how \sys extracts key concepts from the discussion threads.
Next, we explain the mechanism for filtering invalid and redundant rules.
Finally, we present our strategy for managing the dataset size without losing detailed information.
An example of a rule is shown in \autoref{fig:example-rule}.

\para{1. Extracting Concepts.}
\sys extracts key concepts about the patch reviews from the interaction between developers and maintainers.
The interactions contain the vital information that might not be present in existing documentation.
Specifically, this information does not appear in the source code or the Git commit history, as it is only discussed during the code review stage.
We summarize these concepts into a set of rules while also keeping a record of the patch emails for each rule, as the source for the rules might be lost when compressing rules on the concepts.
This is necessary to ensure that the issues reported by our system are reliable and can be traced back to specific old patches from where the rules were generated. 

\para{2. Filtering Invalid and Redundant Rules.}
After extracting the concept and generating the rules, \sys filters out the non-substantive content.
Examples of non-substantive content are vague principles (e.g., use allocator to allocate memory), redundancy (e.g., fix this bug in the next version), and overlapping rules from the same context, as well as communication conventions.
Furthermore, to evaluate the effectiveness of the filter, we randomly sampled 50 rules from the rule set and manually reviewed each one using the following standard: a rule should only guide developers on how to use a specific feature to implement a particular function.
We exclude any rule that does not meet this standard.
As shown in \autoref{tab:rule-stat}, the filtering stage reduced the number of rules by 1.7 times, while achieving a false negative rate of 2\% for invalid rules.

\begin{figure}[t!]
    \centering

\begin{tcolorbox}[
    title={Example Issue},
    colback=black!5!white, 
    colframe=black!75!white, 
    coltitle=white, 
    fonttitle=\bfseries
]
\small

\para{Title: }
Race condition when accessing per-cpu data in preemptible context 

\begin{verbatim}
  static int zswap_frontswap_store(...) {
  	 struct crypto_acomp_ctx *acomp_ctx;
  	 ...
  	 acomp_ctx = raw_cpu_ptr(...);
  	 mutex_lock(acomp_ctx->mutex);
  	 dst = acomp_ctx->dstmem;
  	 ...
  }
\end{verbatim}
> message-id: <current patch message-ID>

\para{Issue Content}

The patch uses `raw\_cpu\_ptr()` to obtain a pointer to the per-cpu `acomp\_ctx` in a preemptible context and then locks a mutex contained within that context. A race condition exists between the `raw\_cpu\_ptr()` call and the subsequent `mutex\_lock()`. If the task is preempted and migrates to a different CPU in this window, `acomp\_ctx` will be a stale pointer to the per-cpu data of the original CPU. The task would then proceed to lock the wrong mutex and use the wrong `dstmem`, leading to data corruption and incorrect synchronization. This same issue exists in `zswap\_frontswap\_load()` and `zswap\_writeback\_entry()`.

\para{Rules Used}

\begin{itemize}
    \item To prevent race conditions from acting on stale state, perform critical checks immediately before the action is taken. If an initial check was performed without a lock, the state must be re-verified under a proper lock before executing the action. Specifically, do not attempt to secure a reference to a kernel object based on an arbitrary address without re-validation, as a race can occur where the memory is freed and reallocated before the reference is taken.
    \subitem > - message-id: <source of the rule>
    \subitem > - Author: <the author of source>
\end{itemize}

\end{tcolorbox}
\caption{A simplified report from about a race condition~\cite{song20:patch-validation-example-bg-src, ahmed25:patch-validation-example-bf-src} reported by \sys.}
\label{fig:example-issue}

\end{figure}

\para{3. Categorizing Rules.}
As we continually generate and collect more rules, the rule set could quickly become so large that it cannot even fit into the LLM's context window when performing patch validation (\autoref{sec:design:validation}).
To solve this problem, \sys reduces the size of the rule set by merging rules based on their topic.
Therefore, \sys follows the conventional human feedback (\autoref{sec:motivation:historical}), categorizing rules into two types: code logic and convention rules, to allow more precise merging.
Code logic concerns system design and security issues, while convention relates to coding style, patch and commit message format, and documentation.

\para{4. Consolidating Rules}
In this stage, we first create a meta rule that covers the same topic of rules, reducing the number of redundant concepts in the rule set. 
We use this method to reduce the number of rules by $15.7$ times, from $3035$ rules to $193$, generated from $177$ discussion threads.
However, this approach can lead to a loss of detailed information if we simply combine the duplicated rules by using LLMs.
Our observations reveal that LLMs may unintentionally merge two different topics of rules when there is only a slight overlap between them.
For example, if one rule addresses control-group concurrency concerns and another discusses memory barriers, LLMs might merge these two rules and overlook the specific details related to one of them.
This occurs because LLMs mistakenly assume that the memory barrier concept is adequately covered within the other concept, which, in reality, it is not.

To avoid information loss, rather than just merging duplicated rules, we tailor the prompts to ensure that only rules on the same topic are merged, making LLMs to preserve all relevant details.
However, even with the prompt restrictions in place, LLMs may still experience some loss of information.
To address this, we introduce a diversity level for each rule, which indicates how often the rule has been referenced in past discussions.
This metric is based on the number of developers (authors) and discussion threads involved.
As shown in \autoref{eq:div-rules}, we quantify the diversity of contributions.
The formula calculates the geometric mean of the number of unique authors, ``\texttt{nr\_author}'', and the number of threads, ``\texttt{nr\_msgid}'', representing the equal balance between two factors with discussion volume.
This ensures that both factors influence the overall diversity score.
Subsequently, we scale the factor of 10 to map the geometric mean to a human-readable integer range and set the threshold at 30 to merge rules with lower diversity levels.
We use this alternative approach to achieve almost $3$ times smaller rule set, while effectively preventing any loss of information.

\begin{equation}
    \label{eq:div-rules}
    diversity\_level = \sqrt{nr\_authors \times nr\_msgid} \times 10
\end{equation}

\begin{table}[t]
    \centering
    \small
    \begin{tabular}{llr}
        \toprule
         \textbf{Stage}            &       \textbf{Rule Category}            &     \textbf{Numbers} \\
         \midrule
         \textbf{1. Extraction}                  & \textbf{}           &       5211   \\
         \midrule
         \textbf{2. Filter}                  & \textbf{}               &       3003   \\
         \midrule
         \multirow{2}{*}{\textbf{3. Categorization}} & Logic         &       1825   \\
                                           & Convention &       1178   \\
         \midrule
         \multirow{2}{*}{\textbf{4. Consolidation}} & Logic         &        592   \\
                                           & Convention   &        437   \\
         \bottomrule
    \end{tabular}
    \caption{Number of rules after each rule generation stage.}
    \label{tab:rule-stat}
\end{table}

\subsection{Patch Validation}
\label{sec:design:validation}

\sys uses the collected rules from \autoref{sec:design:rule-generation} to validate the incoming patch proposals.
In this section, we discuss how \sys uses the rules generated to validate \issues, provide the reliable source for each \issue, and reduce the false \issues.

\begin{figure}[t]
    \centering
    \includegraphics[width=0.925\linewidth]{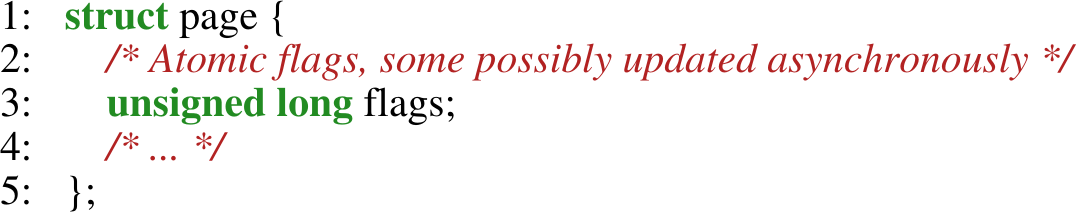}
    \caption{An example of comments providing more details.}
    \label{fig:structure_provide_info}
\end{figure}

\para{1. Retrieve Relevant Source Code}
In this stage, \sys parses the patch content to identify the relevant source code symbols, such as functions and variable types.
By using these symbols, \sys then locates the function call paths and the definition of variable types, and uses this information, like comments, to provide a clear logic and detailed implementation to the LLM, which helps prevent misleading judgments caused by insufficient information.
For example, as shown in \autoref{fig:structure_provide_info}, if the LLM receives a patch proposal that only has partial access to \texttt{page->flags}, it is likely to overlook the data race \issue in this situation due to the lack of detailed information.

\para{2. Retrieve Validation-related Rules}
In this stage, \sys focuses on retrieving the relevant rules to validate patches.
First, \sys instructs the LLM to analyze the patch content and summarize the changes.
Next, \sys examines the source code and iterates through the rule set to prioritize the rules, ranking them from most to least relevant to the patch content.

\para{3. Issue Generation.}
After \sys collects the rules and relevant source code, \sys calls the LLM to review each patch proposal individually, focusing on identifying potential issues within the patch.

\para{4. Batch and Filter.}
After the initial review, \sys batches all the identified issues and filters out possible false positives by analyzing the entire patch set, thereby reducing the number of false alarms.
This is because \kernel development typically follows a step-by-step approach that begins by creating a function prototype and progresses to detailed implementation.
This process aims to transform abstract concepts into specific implementations while minimizing the impact on other subsystems and preventing crashes when individual patches are merged.
Therefore, the issue identified in the first patch may be resolved in a later patch, indicating a false report.

\para{Example.}
\autoref{fig:example-issue} shows an example report from \sys when validating the code in a patch proposal~\cite{song20:patch-validation-example-bg-src}, which has a race condition bug~\cite{ahmed25:patch-validation-example-bf-src}, and provides the issues.
This example shows that \sys can point out the exact bugs, even complex and real-world ones, while also showing the rules used and the references from which the rules were extracted.
This allows developers to easily verify the basis of the report.
\section{Scalable Validation of System Result}
\label{sec:method}

\label{sec:method:challenges}

In the field of machine learning, researchers typically rely on established benchmarks~\cite{jimenez24:swebench, park25:method-for-perf-opt} to evaluate various ML-based tools.
These benchmarks provide a standardized framework for assessment, enabling comparisons across different systems using widely accepted evaluation metrics like precision, F1 score, and recall.
However, a notable gap in the current literature is the absence of studies specifically addressing \kernel patch validation.
Unlike deterministic testing~\cite{zheng23:judging-llma-with-mtbench, du24:bench-class-level-code-gen, lin22:bench-mimic-human-falsehoods}, patch validation systems can generate an unbounded number of potential issues.
This creates a problem space that is impossible to enumerate or fully capture within a single, finite benchmark, as discussed in \autoref{sec:intro}.
Furthermore, manually verifying each reported issue is prohibitively time-consuming, resulting in traditional human-in-the-loop evaluation being unscalable.
Thus, this area requires focused investigation, particularly on developing scalable, reliable, and reproducible validation techniques that minimize human intervention.
Such advancements could significantly enhance the reliability and efficiency of ML-based methods, which are integral to various ML applications.

Evaluating such a system is crucial to help users calibrate the system, in particular, choose the LLM model that best performs within the budget available.
This is important as models vary widely in capability and cost.

Human reviewing~\cite{sun25:llm-code-review, google-sashiko} does not scale and is subjective and prone to inconsistencies.
Moreover, verifying that the generated \issues requires domain-specific knowledge and is challenging and time-consuming, particularly given the unbounded nature of issue discovery.

To address the evaluation problem, this paper proposes a Ground-truth Coverage Score (GCS).
This metric measures how well generated \issues align with the ground-truth of \kernel bug-fix commit messages using past discussions.
It is important to note that the goal of mechanisms is not to replace the developer validation during deployment, ultimately patch proposals need to be validated by developers.
Instead, the goal is to provide an aggregate approximation of the correctness of the system.
As we discuss in \autoref{sec:eval:correctness-of-verifier}, we ultimately confirm the effectiveness of GCS using a manual approach, allowing us to compare the effectiveness of \sys across a vast space of configuration scenarios.

Under the hood, GCS combines rubric-based rewards~\cite{gunjal25:rubrics-rewards-rl}, LLM-as-a-judge~\cite{zheng23:judging-llma-with-mtbench, zhang25:rl-generative-verifiers}, and a Chain-of-Thought~\cite{wei22:chain-of-thought-prompting} (CoT) verifier~\cite{cobbe21:training-verifiers-solve-math, lightman23:lets-verify-step-by-step, wang24:math-shepherd-verify-reinforce-llms}.

\subsection{Ground-truth Coverage Score (GCS)}
\label{sec:method:gcs}

Ground-truth Coverage Score (GCS) implements a backtesting strategy to measure the true positives using past issues found in the kernel code. False positives are inherently harder to measure automatically, thus we measured them manually in our methodology~\autoref{sec:design:validation}.
In practice, GCS measures the performance of the patch validation system by estimating the ground truth coverage of the \issues fixed.

\para{Preliminaries.}
We assume a dataset composed of pairs of patches that were later found to be buggy ($P_b$) with their respective patch fixes ($P_f$). In practice, we use the recent discussions among developers to gather this dataset (\autoref{sec:eval:dataset-collection}).  
A buggy patch ($P_b$) introduces a subsequently fixed problem~\cite{song20:patch-validation-example-bg-src}, while the bug-fix patch ($P_f$) fixes the corresponding issue~\cite{ahmed25:patch-validation-example-bf-src}.
We treat the bug-fix patch ($P_f$) from this dataset as the ground truth ($GT$), given that we aim to automate the existing process.
Finally, we define ``system'' as the candidate that will generate a set of issues for each $P_b$, such as \sys and baseline: 

\begin{equation}
    \begin{gathered}
        I = \{i_x\}_{x=1}^{|I|} = \text{system}(P), \quad \text{for } P \in \{P_b, P_f\}
    \end{gathered}
    \label{eq:sys_gen_issue}
\end{equation}

Later, we utilize a Chain-of-Thought~\cite{wei22:chain-of-thought-prompting} (CoT) reasoning model~\cite{zheng23:judging-llma-with-mtbench, zhang25:rl-generative-verifiers} as the verifier~\cite{cobbe21:training-verifiers-solve-math, lightman23:lets-verify-step-by-step, wang24:math-shepherd-verify-reinforce-llms} for the following judgment framework.
We leverage this model's advanced reasoning capability to evaluate the validity of each \issue.

\para{Evaluation Criteria.}
For each \issue, $i_x$, we use a judgment framework with $P_f$ to generate the \issue-specific criterion~\cite{gunjal25:rubrics-rewards-rl}, denoted as $CP$.
Our framework for evaluation consists of four criteria, where $CP = (c_i)_{i=1}^{4}$, to pose questions for \issue that align with the various properties of the ground truth, $P_f$:
\begin{enumerate}[noitemsep]
    \item \textbf{Root Cause}: Formulate a question about the specific root cause you identified. For example, does the issue describe a problem caused by a race condition between ``function\_A'' and ``function\_B''?
    \item \textbf{Code Location}: Formulate a question about the change surface. For example, does the issue point to code within the patch's change surface, specifically mentioning files like ``file.c'' or functions like ``function\_name''?
    \item \textbf{Fixing Strategy}: Formulate a question about the patch's solution. This is different from the root cause. For example, does the issue suggest a solution that involves ``adding a lock'', ``checking a variable before use'', or a similar strategy implemented in the patch?
    \item \textbf{Keyword or Concept Overlap}: Formulate a question about unique technical terms or concepts from the commit message. For example, does the issue discuss concepts central to the patch description, such as ``a specific memory barrier type'', ``the I/O memory management unit (IOMMU)'', or ``a particular scheduler policy''?
\end{enumerate}

\para{Weighted Confidence Score (WCS).}
To quantify the criteria, the framework uses a CoT-verifier that computes a score for every criterion $c_j$ on a specific issue, $i_x$, as shown in \autoref{eq:wcs}.
This metric integrates the "Yes/No" response with a self-confidence score~\cite{wang23:self-consistency-improves}.
In the formula, $b_{j}^{(P_f,k,n)}(i_x) \in \{0, 1\}$ is a binary correctness function that indicates whether the response "Yes" satisfies the criterion given in the prompt.
And, $c_{j}^{(P_f,k,n)}(i_x) \in [1, 100]$ is a self-reported confidence score~\cite{wang23:self-consistency-improves} from the CoT-verifier.
The framework computes a weighted confidence score based on the positive probability, normalizing the final value to the range $[0, 1]$ to derive the issue's overall score.
\begin{equation}
    \label{eq:wcs}
    \text{WCS}_{j}^{(P_f,k,n)}(i_x) = \frac{1}{100} c_{j}^{(P_f,k,n)}(i_x) \cdot b_{j}^{(P_f,k,n)}(i_x)
\end{equation}

\begin{figure}[t!]
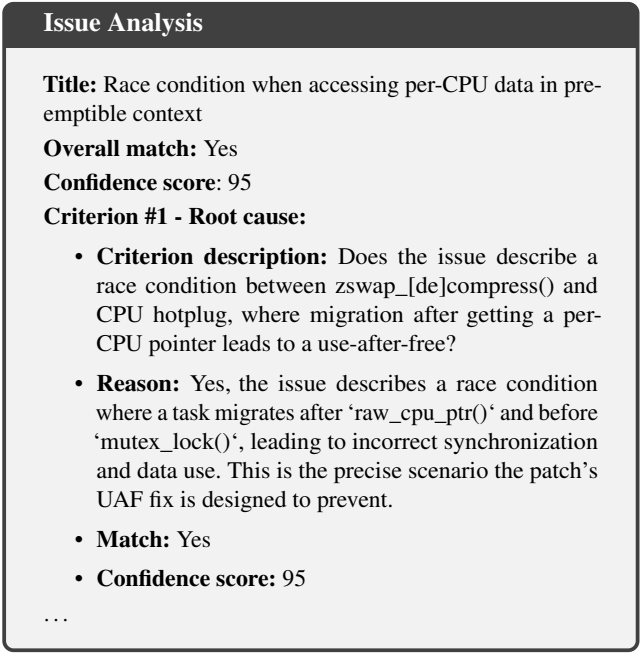

    \centering

\begin{tcolorbox}[
    title={Issue Analysis},
    colback=black!5!white, 
    colframe=black!75!white, 
    coltitle=white, 
    fonttitle=\bfseries
]
\small

\para{Title:}
Race condition when accessing per-CPU data in preemptible context

\para{Overall match:} Yes

\para{Confidence score}: 95

\para{Criterion \#1 - Root cause:}
\begin{itemize}
    \item \textbf{Criterion description:} Does the issue describe a race condition between zswap\_[de]compress() and CPU hotplug, where migration after getting a per-CPU pointer leads to a use-after-free?
    \item \textbf{Reason:} Yes, the issue describes a race condition where a task migrates after `raw\_cpu\_ptr()` and before `mutex\_lock()`, leading to incorrect synchronization and data use. This is the precise scenario the patch's UAF fix is designed to prevent.
    \item \textbf{Match:} Yes
    \item \textbf{Confidence score:} 95
\end{itemize}
$\cdots$
\end{tcolorbox}

    \caption{Abbreviated example of an evaluated issue~\cite{song20:patch-validation-example-bg-src, ahmed25:patch-validation-example-bf-src} by CoT-verifier.
    }
    \label{fig:judge-issue}

\end{figure}

Then, the framework uses the four criteria with WCS.
And, the overall output of CoT-verifier, the CoT-verification rationale, will be the score of the system, which we will discuss further for the consistency of the judgment.
\autoref{eq:cot-verifier-k} shows the entire equation. 
Additionally, \autoref{fig:judge-issue} shows an example of judgment.
\begin{equation}
    \label{eq:cot-verifier-k}
    \text{CoT-verifier}_{(k,n)}(I, P_f) = \frac{1}{|I|}\sum_{x=1}^{|I|} \left( \frac{1}{|CP|} \sum_{j=1}^{|CP|} \text{WCS}_{j}^{(P_f,k,n)}(i_x) \right)
\end{equation}

\para{Ground-truth Coverage Score (GCS).}
To reduce the noise from individual reasoning paths and select the most consistent answer, we employ majority voting~\cite{wang23:self-consistency-improves, zhang25:rl-generative-verifiers}.
Specifically, we generate $K$ CoT-verification rationales and aggregate the `yes' verdicts to calculate an average confidence score.
This aggregation mitigates the impact of reasoning errors from individual CoT-verifiers.
We integrate this voting mechanism into a Best-of-$N$ strategy~\cite{cobbe21:training-verifiers-solve-math, charniak05:coarse-to-fine-n-best}, in which $N$ candidate solutions are generated by the LLM, ranked by the aggregated verification scores, and the top candidate is selected.
Furthermore, we denote the $k$-th verification from majority voting and the $n$-th solution from Best-of-$N$ as $(k, n)$, where $1\leq k \leq K$ and $1 \leq n \leq N$.

Next, we calculate the overall score for every single issue $i_x$ by running the Best-of-N ($N$), majority voting ($K$), and the criterion ($CP$).
We call this the Ground-truth Coverage Score (GCS), reflecting the weighted ground-truth coverage~\cite{lin22:teaching-models2express, zheng23:judging-llma-with-mtbench}, and define it as follows:

\begin{equation}
    \label{eq:ics}
    \begin{gathered}
        GCS_{(\text{system}, N, K)}(P_f, P_b) = \\
        \max_{1 \le n \le N} \left\{ \frac{1}{K} \sum_{k=1}^{K}\text{CoT-verifier}_{(k,n)}(\text{system}(P_b), P_f) \right\}
    \end{gathered}
\end{equation}

Therefore, we can get the system performance from GCS for a single pair of \problem-\solution.
This gives us two key metrics:
(1) ground truth performance: $\text{GCS}_{GT}$.
(2) candidate system performance: $\text{GCS}_{\text{system}}$.
The final verification is the ratio of the candidate's performance to the ground truth's performance:

\begin{equation}
    \label{eq:finalscore}
    \text{FinalScore} = \frac{\text{GCS}_{(\text{system}, N, K)}(P_f, P_b)}{\text{GCS}_{(\text{GT}, N, K)}(P_f, P_f)}
\end{equation}

\para{Highest Ground-truth Coverage Score (H-GCS).}
GCS shows the average quality of the \issues.
To measure the highest coverage, for each CoT-verification rationale, instead of averaging all the issues' scores, H-GCS reports the highest one.

\section{Evaluation}
\label{sec:eval}

\renewcommand{\sys}{\textsc{FLINT}$_{\text{R/S}}$\xspace}

This section first discusses the experiment setup (\autoref{sec:eval:setup} and \autoref{sec:eval:dataset-collection}), and evaluates our system and our methodology along the following questions:

\begin{itemize}[noitemsep]
    \item Can our system detect the issue in \kernel? (\autoref{sec:eval:issue-found})?
    \item What is the accuracy of CoT-verifier with GCS (\autoref{sec:eval:correctness-of-verifier})?
    \item What is the accuracy of issue from our system (\autoref{sec:eval:report-accuracy})?
\end{itemize}

\subsection{Experiment Setup}
\label{sec:eval:setup}

\para{Model.}
We use three open models~\cite{gemma25:gemma3, yang25:qwen3, hui24:qwen2-coder} and two closed models~\cite{deepmind25:gemini2p5} for all the experiments.
The models and their knowledge cut-off are listed in \autoref{tab:model-table}.
We use \texttt{gemini-2.5-flash-preview-09-2025} as \geminitfflash due to the truncated output token problem~\cite{report25:gemini-2-5-flash-truncated-problem}.
However, we still use the stable version of \geminitfflash as CoT-verifier.

\para{Dataset and Knowledge Cut-off.}
To ensure fairness, we split the dataset based on time and the knowledge cut-off point of the latest model, which is January 1, 2025. 
The \problem and \solution data collected before January 1, 2025, are used to generate the rule set.
The \problem and \solution data after this date are used for the validation dataset to prevent data leakage affecting the evaluation.

\para{Kernel Source Code.}
We use Linux \version since its release date, December 29, 2024, is within the cut-off dates for the models that we use.
This allows us to simulate limited information during the review process.

\para{System Setup.}
We evaluate \sys, which uses the rule set and source code, and \llmonly, which does not use the rule set (\autoref{sec:design:rule-generation}) or relevant source code (\autoref{sec:design:validation}), in our evaluation.

\para{CoT-verifier.}
CoT-verifier needs to be at least as powerful as the system's model; thus, it can judge the system result most correctly.
To satisfy this requirement and reduce costs, we use \geminitfpro as CoT-verifier when the system uses \geminitfpro and \geminitfflash, and a stable version of \geminitfflash as CoT-verifier for all the open models.

\begin{table}[]
    \centering
    \small
    \begin{tabular}{lcc}
        \toprule
         \textbf{Model} & \textbf{Knowledge cutoff} & \textbf{Thinking} \\
         \midrule
         \geminitfpro & \multirow{2}{*}{January, 2025~\cite{deepmind25:gemini2p5}} & \multirow{2}{*}{Dynamic} \\
         \geminitfflash$^*$ & \\
         \midrule
         \gemmatfb & \multirow{2}{*}{August, 2024~\cite{gemma25:gemma3}} & \multirow{2}{*}{No}\\
         \gemmatttb & \\
         \midrule
         \qwentttb$^*$ & \makecell{May, 2025~\cite{yang25:qwen3, hui24:qwen2-coder}\\(release date)} & Yes \\
         \bottomrule
    \end{tabular}
    \caption{Information of models in evaluation.
    We use \texttt{qwen3-coder:30b-a3b-q4\_K\_M} as \qwentttb, and \texttt{gemini-2.5-flash-preview-09-2025} as \geminitfflash in the system.
    }
    \label{tab:model-table}
\end{table}

\subsection{Dataset Collection}
\label{sec:eval:dataset-collection}

We construct the dataset using on our analysis of the memory management subsystem from 2024. 
This set covers a wide range of topics from the memory management subsystems, such as synchronization, zswap, process address space, migration, and others.
In total, this data set includes 177 discussion threads and 3934 reply emails.
We use \geminitfpro during the rule collection.
Moreover, we use 21 buggy and bug-fix patch pairs as a validation dataset.

\begin{figure}
    \centering
    \includegraphics[width=0.485\textwidth]{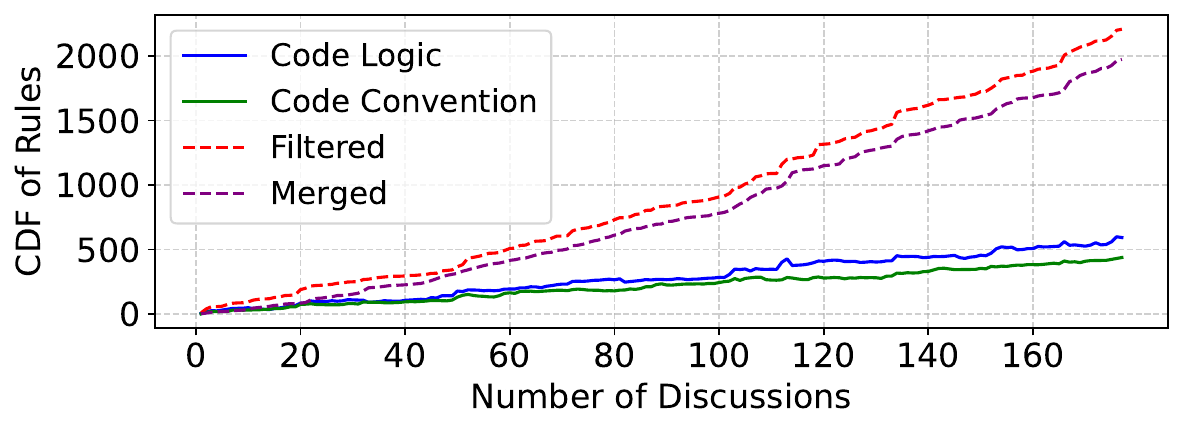}
    \caption{Filtered, merged, and collected code logic and convention rules as threads are progressively analyzed.}
    \label{fig:rule-stats}
\end{figure}

\para{Extracting Rules.}
\sys extracts 1029 rules.
\autoref{fig:rule-stats} shows number of rules across stages as \sys progressively analyzes the 177 discussion threads. 
First, the repeated concept has a positive correlation with the number of discussion threads; as the number of extracted threads increased, the merged rule also increased.
Second, we notice that for each thread, the number of rules has a positive correlation with the number of reply emails. In most of the cases, for the discussions with substantial interactions, the filtered rules have also increased.

\subsection{Issues Found in Linux}
\label{sec:eval:issue-found}

We run \sys on the \lkml to assess the effectiveness of detecting previously unreported issues.
In the v6.18 development cycle, \sys detects 2 new issues in the v6.18 \kernel development, and 7 issues in previous versions.

One of the two new issues identified is a coding convention violation regarding inconsistent function naming.
In particular, \sys detects an inconsistency in a patch proposal that introduced a new feature by adding functions to an existing API family.
While this issue negatively impacts maintainability and readability, it is notoriously difficult for traditional tools to detect.
We reported this issue to the developer, and they confirmed it.
Detecting such issues demonstrates \sys's capability to assist developers in patch reviews.

The second new issue involves variable misuse within the newly introduced code.
As the introduced code is a minor part of the code in \kernel, traditional tools can detect this type of issue, but require significant additional computational resources to do so.
This issue has been reported by the reviewer and confirmed by the developer.

The other 7 issues are either
the design choice,
race condition,
the inconsistent return type with the comment, or
the inconsistency of the commit message description with the implementation.

\subsection{CoT-verifier Accuracy}
\label{sec:eval:correctness-of-verifier}

To evaluate the accuracy of the CoT-verifier's judgments, we conduct a manual review of each criterion associated with any identified issues.
We sample 10 data points for each true or false criterion result from the CoT-verifier and manually analyze the data.
In total, we review 80 criterion results for both the \geminitfpro and \geminitfflash CoT-verifiers.

To evaluate the CoT-verifier against our ground truth, we calculate standard classification metrics including accuracy, precision, recall, specificity, and F1 score.
Additionally, we employ Cohen’s Kappa~\cite{mchugh12:cohens-kappa} to quantify inter-rater reliability and account for chance agreement between the LLM-based CoT-verifiers and humans.
\autoref{tab:verifier-correctness} presents these metrics, highlighting the verifier's performance across different confidence thresholds.

\begin{table}[t]
    \centering
    \small
    \renewcommand{\arraystretch}{1.1} 
    \begin{tabular}{l|c|cc}
        \toprule
        & \textbf{\geminitfpro} & \multicolumn{2}{c}{\textbf{\geminitfflash}} \\
        \cmidrule(lr){2-2} \cmidrule(lr){3-4}
        \textbf{Metric} & \textbf{Full Set} & \textbf{Full Set} & \textbf{High Conf.} \\
         & {\scriptsize ($N=80$)} & {\scriptsize ($N=80$)} & {\scriptsize ($>90$, $N=41$)} \\
        \midrule
        \textbf{Cohen's Kappa} & 0.925 & 0.475 & 0.809 \\
        \textbf{Precision}     & 0.950 & 0.500 & 0.750 \\
        \textbf{Recall}        & 0.974 & 0.952 & 1.000 \\
        \textbf{F1 Score}      & 0.962 & 0.658 & 0.857 \\
        \textbf{Specificity}   & 0.951 & 0.661 & 0.906 \\
        \bottomrule
    \end{tabular}
    \caption{Performance and inter-rater reliability of the CoT-verifier against human review. The evaluation covers 80 judgments balanced across correctness labels (10 true and 10 false per criterion). We compare \geminitfpro (verifying \geminitfpro) against \geminitfflash (verifying \gemmatfb). \textbf{Full Set} includes all 80 judgments; \textbf{High Conf.} refers to the subset where the confidence score is $>90$.}
    \label{tab:verifier-correctness}
\end{table}

\para{\geminitfpro CoT-verifier.}
The overall agreement accuracy is 96.25\%.
The precision, recall, specificity, and F1 score further demonstrate that the CoT-verifier is generally accurate.
Specifically, out of 80 total cases, there were only 3 instances of disagreement between the CoT-verifier and the reviewer: 2 false positives and 1 false negative relative to the reviewer.
Furthermore, Cohen's Kappa indicates an almost perfect agreement between the CoT-verifier and human reviewers, with a value of $\kappa = 0.925$ and a 95\% confidence interval of $[0.842, 1.000]$.

\para{\geminitfflash CoT-verifier.}
The overall agreement accuracy is 73.75\%.
The metrics in \autoref{tab:verifier-correctness} show lower accuracy than \geminitfpro.
The confusion matrix shows that \geminitfflash has a systematic bias, which has high sensitivity ($0.95$) and low precision ($0.50$).
This discrepancy (20 false positives and 1 false negative) suggests that \geminitfflash applied a significantly more relaxed threshold than \geminitfpro.
Also, Cohen's Kappa indicates moderate agreement ($\kappa = 0.475$), and 95\% confidence interval $[0.282, 0.668]$, which has the lowest level of agreement with a human reviewer.

\para{\geminitfflash CoT-verifier With High Confidence.}
To enhance the reliability of \geminitfflash on the CoT-verifier, we further analyze the judgments and confidence scores it provides.
We investigate the incorrect criterion judgments to evaluate the average confidence scores for both incorrect and correct judgments.
The average confidence score for incorrect judgments is $80.24$, while for correct judgments, it is $92.63$.
This indicates that with \geminitfflash, if we disregard confidence scores below $90$, we are likely to achieve results that closely align with a human reviewer.
Based on this observation, we set the threshold of the weighted confidence score to obtain a more accurate score using \geminitfflash.

Applying a confidence threshold of 90 significantly enhances the reliability of the lightweight model.
Specifically, filtering for high-confidence judgments raises the Cohen's Kappa for \geminitfflash to substantial agreement ($\kappa = 0.809$), with a 95\% confidence interval of $[0.601, 1.000]$, which is close to almost perfect agreement.
This confirms that \geminitfflash is capable of high-fidelity verification when the evaluation is restricted to its most confident predictions.

In summary, by breaking down judgment into smaller, specific criteria, \geminitfpro performs similarly to human judgment.
In contrast, while \geminitfflash can identify when the results do not align with the ground truth, \geminitfflash struggles with accurately assessing when a result is close to the ground truth if the judgment's confidence score is below 90.
However, by filtering out judgments with a confidence score below 90, \geminitfflash can perform close to human.

\subsection{Report Accuracy}
\label{sec:eval:report-accuracy}

\begin{table}[t] 
    \centering
    \small
    \addtolength{\tabcolsep}{-2pt}
    \renewcommand{\arraystretch}{0.75}
    \begin{tabular}{cccc|ccc}

\toprule
%
%
\textbf{Score} &  \textbf{GA3:4b} & \textbf{GA3:12b} & \textbf{Q3C:30b} & \textbf{G2.5F} & \textbf{G2.5P} \\
\midrule
\multirow{2}{*}{\textbf{GCS}} & \cellcolor{lightgray} 0.18 & \cellcolor{lightgray} 0.35 & \cellcolor{lightgray} 0.51 & \cellcolor{lightgray} 0.63 & \cellcolor{lightgray} 0.65 \\
& 0.43 & 0.41 & 0.47 & 0.71 & 0.63 \\
\multirow{2}{*}{\textbf{H-GCS}} & \cellcolor{lightgray} 0.27 & \cellcolor{lightgray} 0.40 & \cellcolor{lightgray} 0.75 & \cellcolor{lightgray} 0.72 & \cellcolor{lightgray} 0.67 \\
&      0.52 &      0.59 &      0.78 &      0.77 &      0.75 \\

        \bottomrule
    \end{tabular}
    \caption{The true positives analysis.
            Overall GCS and H-GCS with Best-of-$10$ and majority voting, $K=3$.
            \colorbox{lightgray}{\sys} is gray color, and \llmonly is white color.
            \geminitfpro (G2.5P) and \geminitfflash (G2.5F)'s CoT-verifier are \geminitfpro, others, Gemma3 (GA3:4b and GA3:12b) and \qwentttb (Q3C:30b), CoT-verifier is \geminitfflash. Higher is better.}
    \label{tab:gcs}
\end{table}

We evaluate the capabilities of \sys, which reports \issues based on the rule set, using the GCS metric described in \autoref{sec:method:gcs}.
We use the validation dataset, where each data point consists of a \problem and \solution pair, and all data points are after January 1, 2025.
We run \sys and \llmonly with 5 different models.
For the open models, we use \geminitfflash as the CoT-verifier, while for the closed models, we use \geminitfpro.
The \solution is considered the ground truth in this evaluation.
Among the validation dataset, only one data point’s ground truth received a score of less than the maximum (1) on the GCS metric, which is buffer overflow.
For every data point, we run each model and system 10 times to obtain the Best-of-$10$ outcomes using $K=3$ majority voting.
We categorize all the data points into 8 types,
coding convention (\textbf{style}),
uninitialized local variable (\textbf{ULV}),
design choice and implementation error (\textbf{design}),
performance regression (\textbf{perf}),
locking issues (\textbf{lock}),
data race and race condition (\textbf{racing}),
use-after-free (\textbf{UAF}),
and buffer overflow (\textbf{BOF}),
to illustrate the different capabilities on different bug patterns.

\para{Ground-truth Coverage Score (GCS).}
We first evaluate the GCS on both systems to show the average performance across all the issues and trials.
A high GCS indicates that the system can consistently identify the ground truth.
In other words, the system is able to consistently pinpoint the true bug we specify.
As shown in \autoref{tab:gcs}, \sys achieves an average improvement 8\% with \qwentttb and 3\% with \geminitfpro.
However, \sys does not have the same performance on weaker models, indicating that the stronger reasoning or coding-specific models are more suitable for the patch validation.

\begin{figure}
    \centering
    \begin{subfigure}[t]{0.445\textwidth}
        \includegraphics[width=\linewidth]{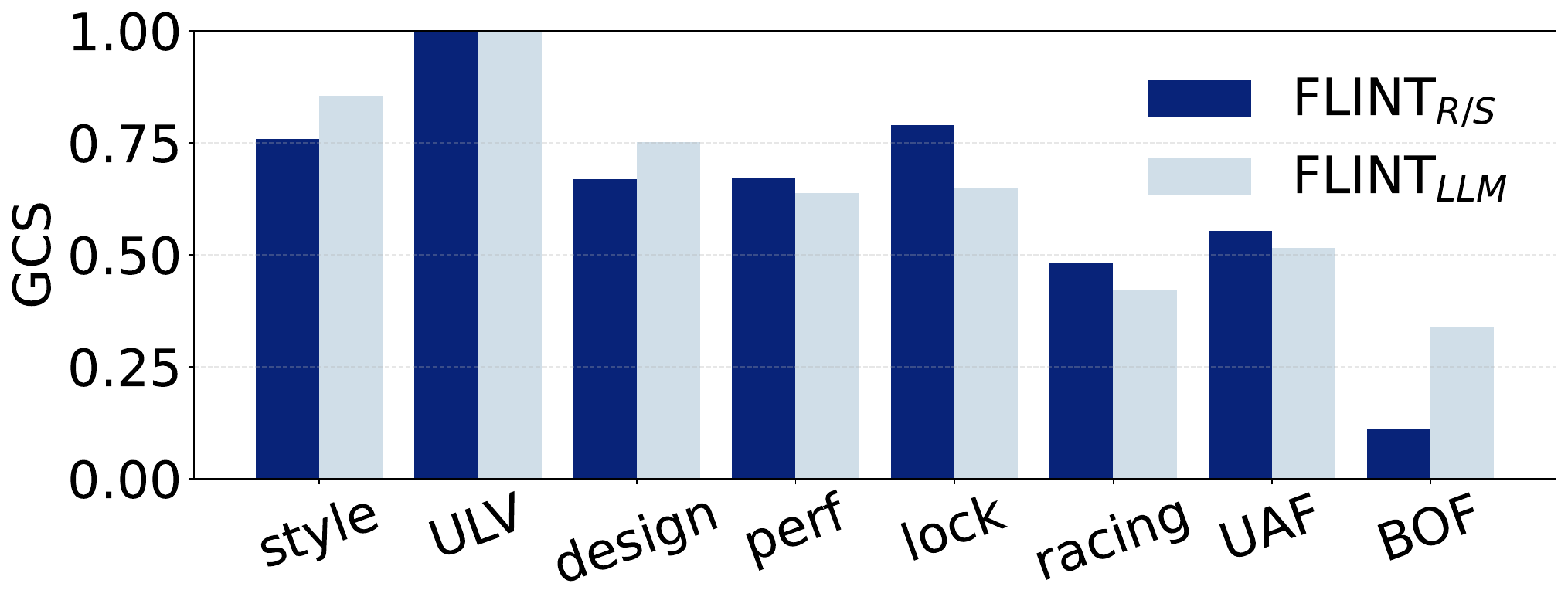}
        \caption{GCS of \geminitfpro.}
        \label{subfig:gcs_gemini-2-5-pro}
    \end{subfigure}
    \begin{subfigure}[t]{0.445\textwidth}
        \includegraphics[width=\linewidth]{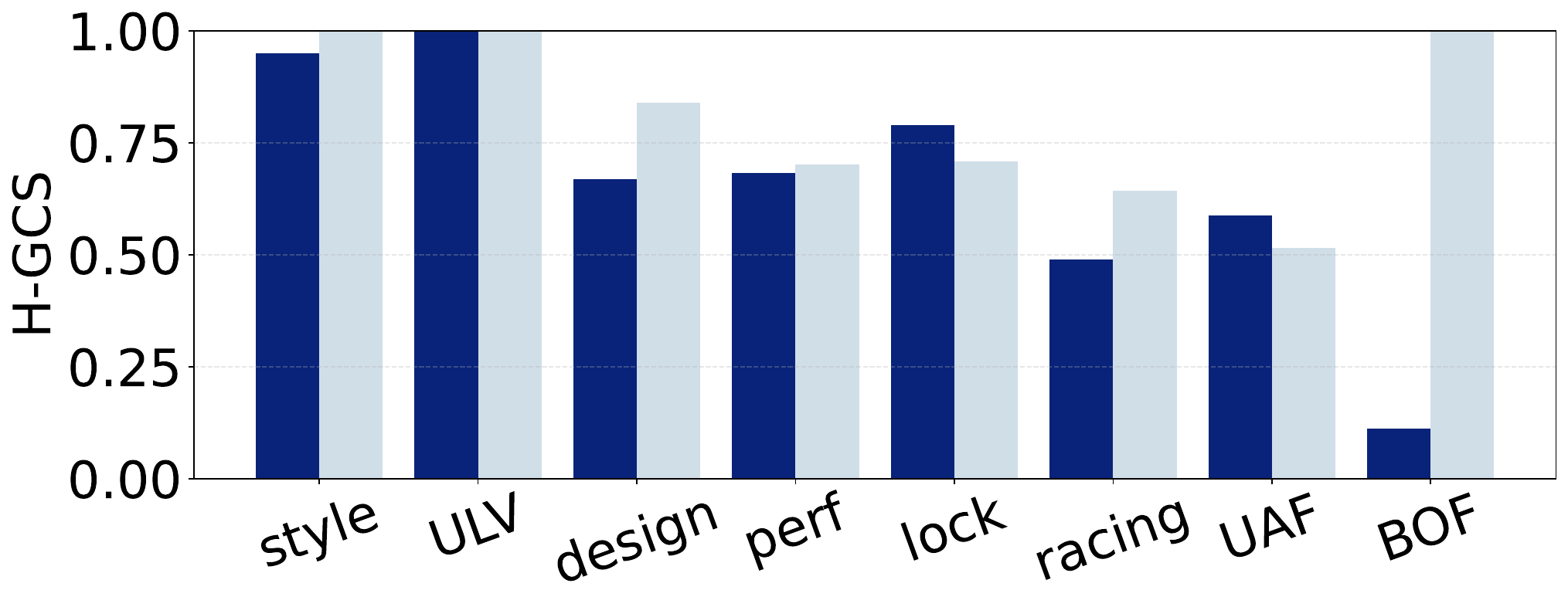}
        \caption{H-GCS of \geminitfpro.}
        \label{subfig:hgcs_gemini-2-5-pro}
    \end{subfigure}
    \caption{The true positives analysis. GCS and H-GCS with Best-of-10 and majority voting, $K=3$, of \geminitfpro.
    }
    \label{fig:all_gcs_gemini-2-5-pro}
\end{figure}

We further detailed analyze \sys on each type of issue.
\autoref{subfig:gcs_gemini-2-5-pro} shows that with the strongest model, \geminitfpro \sys achieves higher scores on lock (21\%), racing (14\%), perf (4\%), and UAF (6\%).
On the other hand, \llmonly performs better on style, design, and buffer overflow.
This difference primarily comes from the rule set used by \sys, which is collected from unbalanced past discussions on concurrency topics.
In contrast, \llmonly is not restricted by the rule set.
Therefore, across 10 trials, \llmonly can generate a more diverse range of issues, allowing it to cover the ground truth more effectively.

\para{Highest Ground-Truth Coverage Score (H-GCS).}
We then evaluate the H-GCS to show the peak ground truth coverage across trials.
A high H-GCS indicates that, among all the issues in a trial, the system can accurately pinpoint the specific true bug.
As shown in \autoref{tab:gcs}, \sys generally underperforms compared to \llmonly.
This is primarily because \sys is constrained to reporting the \issue only based on the rule set.
However, looking into specific types in \autoref{subfig:hgcs_gemini-2-5-pro} with \geminitfpro, \sys achieves better scores on lock (11\%) and UAF (13\%).
Furthermore, when comparing the GCS and H-GCS for \sys, we notice minimal differences.
In contrast, \llmonly shows significant variation in some types, such as data race, buffer overflow, and design/logic issue.
For example, \llmonly has a GCS of $0.34$ and H-GCS of $1.00$ on BOF type with \geminitfpro.
This difference is because \sys provides more consistent issues across trials based on the reliable rule set, whereas \llmonly may experience more hallucinations, leading to a higher false positive rate (\autoref{sec:eval:fpr}).

\para{False Positives.}
\label{sec:eval:fpr}
In this experiment, we evaluate the false positive rate of \sys and  \llmonly. To determine the ground truth for each issue, we manually check to see if a sample of issues reported by our system have been fixed in the recent version 6.18 of the \kernel and/or discussed in the \lkml. 
Thus, this is a conservative analysis, as some issues might be true positives but not yet identified by developers.
Due to the substantial verification workload involved in analyzing each \issue, we randomly sampled 20 \issues from \sys and \llmonly.
Both \sys and \llmonly use \geminitfpro in this experiment.

\begin{table}[t]
    \centering
    \small
    \begin{tabular}{lccccc}
        \toprule
        \textbf{System} & \textbf{Total} & \textbf{TP} & \textbf{FP} & \textbf{FPR} & \textbf{Prec.} \\
        \midrule
        \sys     & 20 & 13 & 7  & 35\% & 65\% \\
        \llmonly & 20 & 9  & 11 & 55\% & 45\% \\
        \bottomrule
    \end{tabular}
    \caption{False report rate and precision of \sys versus \llmonly with \geminitfpro.}
    \label{tab:fp_issues}
\end{table}

As shown in \autoref{tab:fp_issues}, \sys has a precision of 65\% and a false positive rate of 35\%, which is nearly twice as low as that of \llmonly, which has a precision of 55\%.
This indicates that \sys provides more accurate results with fewer false reports compared to \llmonly.
Additionally, most of the true positive \issues have already been fixed in the most recent version.

\renewcommand{\sys}{\textsc{FLINT}\xspace}
\section{Related Work}
\label{sec:related-work}

\para{Automated Code Review.}
Several works~\cite{sun25:llm-code-review, google-sashiko} utilize LLMs to conduct code reviews.
BitsAI-CR~\cite{sun25:llm-code-review} employs a customized fine-tuned LLM to automate patch review, while adopting costly human annotation to collect developer feedback, creating a flywheel mechanism to continuously improve BitsAI-CR.
Sashiko~\cite{google-sashiko} uses an agentic code review system with a set of Linux kernel-specific prompts and protocol to automatically review.
In contrast, \sys utilizes a rule set derived from past discussions to sustain and enhance its accuracy without the need for costly fine-tuning, human annotation, or a heavy agentic system.

\para{Static Analysis.}
The static analysis tools for \kernel, such as Sparse~\cite{sparse-doc} and Smatch~\cite{smatch}, use a semantic parser to inspect and report on the abstract syntax tree of a C program with kernel-specific annotation.
These tools usually depend on the manual and the specific annotations to reduce the false positive rate~\cite{smatch, ryan24:data-race-prediction-in-linux}.
In comparison, \sys has a lower false positive rate and can detect not only semantic errors but also design and coding convention issues.

\para{LLM-Synthesized Static Analysis.}
An alternative approach~\cite{wu25:llm-bug-pattern-finding, yang25:knighter-transforming-static-analysis} to preventing security vulnerabilities is to use a machine learning-based method to identify similar bug patterns from past data.
For example, BugStone~\cite{wu25:llm-bug-pattern-finding} employs this technique. 
BugStone introduces a rule-centric system that analyzes recurring bug patterns from past research to identify similar bugs in source code.
Additionally, Knighter~\cite{yang25:knighter-transforming-static-analysis} utilizes LLMs to create a set of static analysis tools for direct analysis of the source code.
These LLM-based static analysis tools are close to \sys but have several differences.
First, \sys aims to find the bug with the fragmented information of the source code.
Second, \sys can detect the design and coding conventions issues in the patch proposals.
On the other hand, these LLM-based static analysis tools need the full source code and focus on finding existing logic bugs.

\para{Testing Tools.}
The traditional~\cite{syzkaller, syzbot, intel-lkp} and ML-based testing tools~\cite{gong25:ml-fuzz-snowplow, gong23:ml-fuzz-snowcat, wang21:ml-fuzz-syzvegas, yang25:ml-fuzz-kernelgpt} for \kernel usually tend to take more than an hour to find the bug, which is not a suitable case for rapid interactive response in \kernel development.
For example, the Linux Kernel Performance~\cite{intel-lkp} (LKP) only tests specific kernel versions weekly, and Syzbot~\cite{syzbot, syzkaller} also focuses on specific kernel versions.
This coarse and relatively long-term testing makes it difficult to provide immediate feedback and limits the testing to accepted patches, which does not alleviate the workload associated with patch reviews.

\para{LLM-Synthesized Testing.}
Several studies~\cite{liu25:llm-generate-regression-tests, mathai24:kgym-platform-dataset-benchmark} focus on generating test input to validate source code.
For example, Cleverest~\cite{liu25:llm-generate-regression-tests} is a feedback-directed, zero-shot LLM-based regression test generation technique that can create test cases for new patches.
Additionally, KGym~\cite{mathai24:kgym-platform-dataset-benchmark} provides a virtual environment for testing and patching kernels to detect bugs.
These studies are independent and orthogonal to our system, as they aim to perform runtime tests on software.
This approach requires more computational resources and is not designed to handle the high throughput of incoming patches in the kernel.
\section{Conclusion}
\label{sec:conclusion}

This paper presents a 10 years study of the patch review process in the \kernel memory subsystem.
To address the problems identified in the study, this paper introduces a novel approach called \sys, a rule-based patch validation system.
\sys effectively bridges the gap between the reasoning capabilities of LLMs and the practical constraints of the patch review process.
\sys achieves $1.6\times$ lower false positive rate, and detects 2 issues in the recent \kernel version.

\bibliographystyle{plain}
\bibliography{bib/patch_validation, bib/related, bib/llmjudge, bib/arxiv}

\end{document}